\documentclass[prd,aps,twocolumn,amsmath,amssymb,nofootinbib,preprintnumbers]
{revtex4}

\usepackage{graphicx}
\usepackage{dcolumn}
\usepackage{bm}
\usepackage{amsmath}
\usepackage{amsthm}

\usepackage{color}

\usepackage{amsfonts}
\usepackage{euscript,bbm}
\usepackage{ifthen}
\usepackage{psfrag}
\usepackage{slashed}

\def\ls{\mathrel{\lower4pt\vbox{\lineskip=0pt\baselineskip=0pt
           \hbox{$<$}\hbox{$\sim$}}}}
\def\gs{\mathrel{\lower4pt\vbox{\lineskip=0pt\baselineskip=0pt
           \hbox{$>$}\hbox{$\sim$}}}}
\def\drawbox#1#2{\hrule height#2pt

\hbox{\vrule width#2pt height#1pt \kern#1pt
              \vrule width#2pt}
              \hrule height#2pt}

\def\Asym#1#2{\vcenter{\vbox{\drawbox{#1}{#2}
              \kern-#2pt       
              \drawbox{#1}{#2}}}}

\newcommand{\be}{\begin{equation}}
\newcommand{\ee}{\end{equation}}
\newcommand{\bea}{\begin{eqnarray}}
\newcommand{\eea}{\end{eqnarray}}

\newcommand{\neu}[1]{\ensuremath{\tilde{\chi}_{#1}^0}}

\newcommand{\gsim}{\lower.7ex\hbox{$\;\stackrel{\textstyle>}{\sim}\;$}}
\newcommand{\lsim}{\lower.7ex\hbox{$\;\stackrel{\textstyle<}{\sim}\;$}}

\newcommand{\ttbar}{t \bar{t}}

\newcommand{\met} {{E\!\!\!\!/_{\rm T}}}
\newcommand{\mht} {{H\!\!\!\!/_{\rm T}}}

\newcommand{\pT} {{p_{\rm T}}}

\newcommand{ \pythia } {{\tt PYTHIA}}
\newcommand{ \delphes } {{\tt DELPHES}}

\newcommand{ \madgraph } {{\tt MADGRAPH5}}

\begin{document}
MI-TH-1519
%
\title{Probing Compressed Bottom Squarks with Boosted Jets and Shape Analysis}


\author{Bhaskar Dutta$^{1}$, Alfredo Gurrola$^{2}$,Kenichi Hatakeyama$^{3}$, Will Johns$^{2}$, Teruki Kamon$^{1,4}$, Paul Sheldon$^{2}$, Kuver Sinha$^{5}$, Sean Wu$^{1}$, Zhenbin Wu$^{3,6}$}

\affiliation{$^{1}$~Mitchell Institute for Fundamental Physics and Astronomy, \\
Department of Physics and Astronomy, Texas A\&M University, College Station, TX 77843-4242, USA \\
$^{2}$~Department of Physics and Astronomy, Vanderbilt University, Nashville, TN, 37235, USA \\
$^{3}$~Department of Physics, Baylor University, Waco, TX 76798-7316, USA\\
$^{4}$~Department of Physics, Kyungpook National University, Daegu 702-701, South Korea \\
$^{5}$~Department of Physics, Syracuse University, Syracuse, NY 13244, USA  \\
$^{6}$~Department of Physics, University of Illinois at Chicago, Chicago, IL 60607-7059, USA
}

\begin{abstract}

A feasibility study is presented for the search of the lightest bottom squark
(sbottom)
in a compressed scenario, where its mass difference from the lightest
neutralino is 5 GeV. Two separate studies are performed: $(1)$ final state
containing two VBF-like tagging jets, missing transverse energy, and zero or
one $b$-tagged jet; and $(2)$ final state consisting  of initial state
radiation (ISR) jet, missing transverse energy, and at least one $b$-tagged jet. An
analysis of the shape of the missing transverse energy distribution for signal
and background is performed in each case, leading to significant improvement
over a cut and count analysis, especially after incorporating the
consideration of systematics and pileup. The shape analysis in the VBF-like tagging jet study
leads to a $3\sigma$ exclusion potential of sbottoms with mass up to $530 \,
(462)$  GeV for an integrated luminosity of $300$ fb$^{-1}$ at 14 TeV, with
$5\%$ systematics and PU $= 0 \, (50)$.



\end{abstract}
\maketitle


{\it {\bf Introduction - }}  Weak-scale supersymmetry addresses the hierarchy problem, gives gauge coupling unification, and (in $R$-parity conserving models) provides a robust dark
matter (DM) candidate, the lightest neutralino ($\neu{1}$). As such, it is one of the most widely studied frameworks for physics beyond the Standard Model (SM).

The exclusion bounds on supersymmetric colored particles belonging to the
first two generations are already quite strong. For comparable squark
($\tilde{q}$) and gluino ($\tilde{g}$) masses, the data eliminates these
particles up to approximately $1.5$ TeV at $95\%$ C.L. with $20$ fb$^{-1}$ of
integrated luminosity \cite{:2012rz, Aad:2012hm, CMS:2014nia, CMS:2012mfa,
LHCsquarkgluino20ifb}. 

On the other hand, the bounds on the masses of the colored third generation
are much weaker due to smaller production cross section. Given that a new
boson consistent with the SM-like Higgs has been observed, with mass in the
region of $125$~GeV \cite{LHCHiggs}, a weakly coupled light scalar must have
its mass stabilized against quantum corrections. Probing top squarks (stops)
is a high-priority study for the future given its importance in this context. 

Since left handed bottom squark (sbottoms, $\tilde{b}$) and stops come in the
same electroweak doublet, it is equally important to search for light sbottoms. Moreover, light sbottoms
can play a role in obtaining the correct relic density of a neutralino DM
candidate, through coannihilation effects \cite{Griest:1990kh}. Sbottom pairs produced from QCD interactions will decay to the lightest superpartner, which we will assume to
be a stable neutralino, through the process $\tilde{b} \rightarrow b \neu{1}$.
When the mass difference between the sbottom and the neutralino is large, the
$b$-tagged jet is sufficiently boosted. In that case, the standard procedure for the
sbottom search involves final states containing two jets, at least one of
which is $b$-tagged, and large missing transverse energy $\met$ coming from the
neutralino. This has been the strategy of sbottom searches at CMS and ATLAS. 

The current exclusion bounds on sbottoms are as follows. With $20$ fb$^{-1}$ of
data at $8$ TeV, the ATLAS Collaboration has ruled out sbottoms up to $650$
GeV, for neutralino masses less than $300$ GeV \cite{Aad:2013ija}, and up to
$255$ GeV for the mass-degenerate scenario~\cite{Aad:2014nra}.  Similar
exclusion bounds have been obtained by the CMS Collaboration when the similar
search strategy is employed~\cite{CMS:2014nia, Khachatryan:2015wza}.


The purpose of this paper is to propose a new search strategy for sbottoms in
compressed regions of parameter space, where the $m_{\tilde{b}} - m_{\neu{1}}$
mass difference is small (we have kept this value at 5 GeV throughout this study). The challenge in the compressed region is that both $\met$ and the $p_{\rm T}$ of the $b$-tagged jets become small, due to insufficient boosting of the objects coming from the sbottom decay. We point out that this can be overcome in final state topologies containing two boosted forward jets
in opposite hemispheres (reminiscent of vector boson fusion jets). Gluons
radiated off of the forward jets can produce sbottoms which decay to $b$-tagged jets
and $\met$ in the central region of the detector. To balance the high initial
$p_{\rm T}$ of the incoming partons, the centrally produced decay products are
boosted, even in the compressed region.

This topology has been proposed by some of the authors as a probe of
both colored and non-colored supersymmetric particles. Charged and neutral
Wino production followed by decays to $\neu{1}$ via a light slepton has been
studied in \cite{Dutta:2012xe,cho}, while searches for Winos and Higgsinos in
the final state of two VBF-tagged jets and $\met$ has been proposed in
\cite{Delannoy:2013ata}. Moreover, top squarks have been studied in the
compressed region using this topology \cite{Dutta:2013gga}.

The search strategy using the initial state radiation (ISR) jet has been
already employed by the CMS and ATLAS Collaborations, and has shown a good
sensitivity for signals with compressed spectra~\cite{CMS:2014nia,
Aad:2014nra}. We also consider a separate study with final state consisting
of an ISR jet, $\met$, and at least one $b$-tagged jet.

In each case, the shapes of the $\met$ distributions for signal and background
are studied using a binned likelihood following the test statistic based on
the profile likelihood ratio. We find that shape analysis, compared to a
simple cut and count analysis, yields a significant improvement of the
compressed sbottom mass reach after incorporating the effects of systematics
and pileup.

In the rest of the paper, we present results first from the analysis with VBF-like tagged jets in the final state, and next the analysis with a ISR jet in the final state. We end with our discussions.

{\it {\bf VBF-like Tagging Jets Study -}} For this feasibility study, inclusive $\tilde{b} \tilde{b}^{*} \, + \,$ multijets samples are generated with $\tilde{b}$ masses in the range of $15$--$1000$ GeV, keeping $m_{\tilde{b}} - m_{\neu{1}} \, \sim \, 5$ GeV.  Both QCD and
weak production processes are included. The $\neu{1}$ in our studies is mostly
Bino and the sbottom decays entirely through the canonical
channel $\tilde{b} \rightarrow b \neu{1}$. The other colored particles,
neutralinos and charginos are assumed to be much heavier.

Signal and background samples are generated with \madgraph \,
\cite{Alwall:2011uj} followed by the parton showering and hadronization with
\pythia \, \cite{Sjostrand:2006za} and the detector simulation using \delphes \,
\cite{deFavereau:2013fsa}. One advantage of the \delphes \, simulation is that it
can simulate pileup $pp$ interactions, which was not possible with other fast
simulation programs available in the HEP community. We used the Snowmass
detector configuration as defined in \cite{Anderson:2013kxz}, which represents the typical
performance of the CMS and ATLAS detectors.
We perform a $14$ TeV study for pileup PU $ = 0, \, 50,$ and $140$ scenarios,
with an assumption of 5\% systematics on the signal and background.

Given the compressed spectrum, we explored the sensitivity with two VBF-like
tagged jets and large missing transverse energy final state. We also found
that with the VBF-like event signature, the $b$ jets from the sbottom decays
also get boosted in the central part of the detector, providing another
approach for probing this compressed signal. Overall, two studies are
performed: $(1)$ final state with two VBF-like tagged jets,  $\met$ and zero
$b$-tagged jet and $(2)$ final state with two VBF-like tagged jets, $\met$,
and one $b$-tagged jet.

The cuts employed are as follows.

$(1)$ $\mht$-$\met$ asymmetry cut: the condition
\be
\frac{|\mht - \met|}{\mht + \met} \, < \, 0.2 \, (0.5) \,\,\, {\rm for} \,\, {\rm PU} = 0 (140)
\ee
is imposed to protect against occasional loss of high $p_{\rm T}$ jets due to
the aggressive pileup subtraction in \delphes \,. Here, $\mht$ is defined as
the negative vectorial sum of jets with $\pT \geq 30 $ GeV, muons, electrons,
and photons. The $\mht$ was found to less pileup dependant, and used instead
of $\met$ in this analysis.

$(2)$ Boosted jet cuts: the event is required to have the presence of at least
two jets ($j_1$, $j_2$) satisfying: $(i)$ $\pT(j_1)  \geq 50 (200) $ GeV and
$\pT(j_2)  \geq 50 (100) $ GeV for PU $ = 0 \, (140)$ in $|\eta| \leq 5$;
$(ii)$ $|\Delta \eta (j_1, j_2)| > 4.2$; $(iii)$ $\eta_{j_1} \eta_{j_2} < 0$;
$(iv)$ dijet invariant mass $M_{j_1 j_2} \, > \, 1500$ GeV;
$(v)$ missing transverse energy $\mht \, > \, 50$ GeV.

$(3)$ Vetoes: We veto electrons, muons and tau-tagged jets. In the final state
study with two VBF-like tagged jets,  $\met$ and zero $b$-tagged jet, a $b$-tagged jet
veto is also applied at this stage. 

$(4)$ $b$-tagged jet requirements (for $1 \, b$-tagged jet study):  For the
tagged jets plus $\met$ plus $b$ study, we require exactly one tight
$b$-tagged jet in the final state, with $p_{\rm T} \, < \, 80$ GeV to suppress
the $\ttbar$ background. 


$(5)$ $\mht$ cut: We require $\mht > 200$ GeV.

$(6)$ $\Delta \phi_{jj} $: From the search for invisible Higgs boson in the
vector boson fusion at CMS~\cite{Chatrchyan:2014tja}, the QCD background is
reduced to a low level by requiring the azimuthal separation between the
VBF-tagged jets to be small. Here, we require $\Delta \phi_{jj} $ $<$ 1.8.

The cut flow table for the benchmark point for PU $=0$ in the VBF-like tagged
jets, $\mht$, and zero $b$-tagged jet study is presented in Table
\ref{tablesbenchmark}.

\begin{table}[!htp] 
\caption{Cut flow table for $m_{\tilde{b}} = 500$ GeV with PU $= 0$, in the
final state with two VBF-like tagged jets, $\mht$, and zero $b$-tagged jets.}
\label{tablesbenchmark}
\begin{center}
\begin{tabular}{c c c }     
    \hline \hline \\

    Selection \,\, & Signal (pb) \,\, & Background (pb)  \\

    \hline \hline \\
    
    Boosted jets \,\, &  5.6 $\cdot 10^{-3}$ \,\, &  10.2   \\
    $b$-tagged jet veto \,\, &  5.5 $\cdot 10^{-3}$ \,\, &  9.8   \\
    Lepton veto \,\, &  5.4 $\cdot 10^{-3}$ \,\, &  6.5 \\
    $\mht \, > \, 200$ \,\, &  3.3 $\cdot 10^{-3}$ \,\, &  0.6  \\
    $\Delta \phi$ \,\, &  2.1 $\cdot 10^{-3}$ \,\, &  0.25 \\

    \hline \hline

\end{tabular}
\end{center}
\end{table}

\begin{figure}[!htp]
\centering
\includegraphics[width=3.5in]{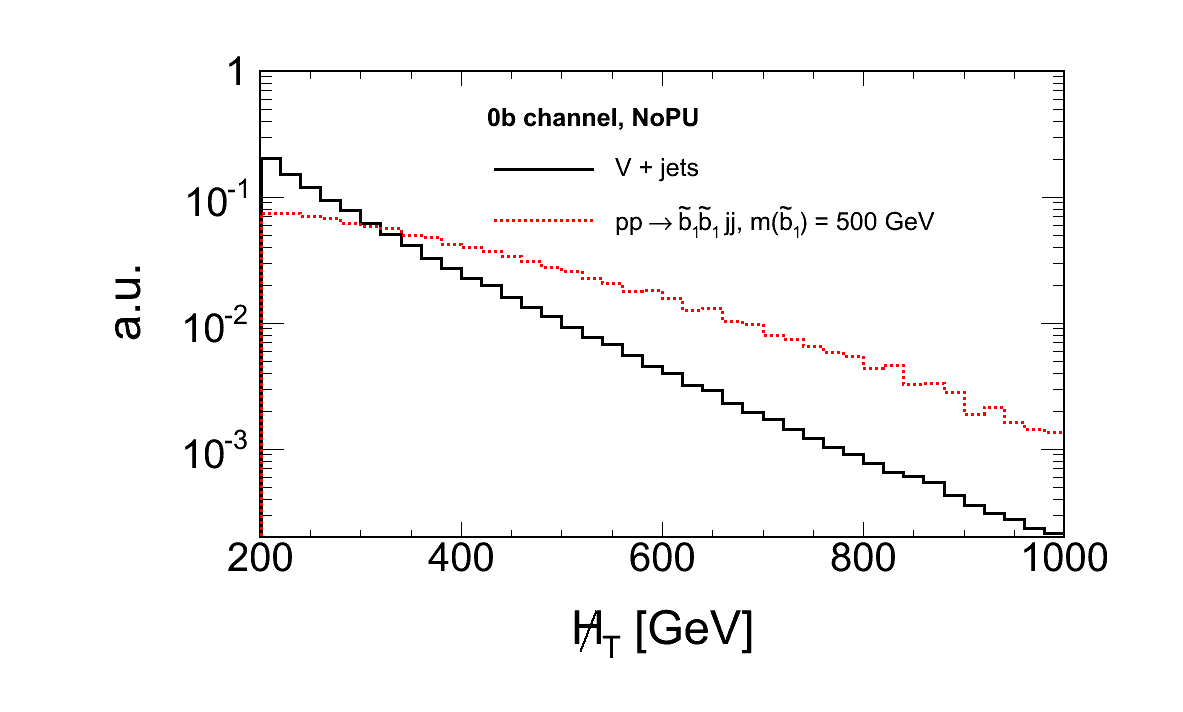}
\caption{Distributions of $\mht$ normalized to unity for signal (green dotted
histogram) and dominant $V + $jets background (black solid histogram) after
all selections except $\mht$ requirement, for the benchmark point with
$m_{\tilde{b}} = 500$ GeV, $m_{\neu{1}} = 495$ GeV, in the case of PU $=0$,
for the channel with VBF-like tagged jets, $\mht$, and zero $b$-tagged jet.}
\label{MHT_BGvsSignal_NoPU}
\end{figure}

Figure \ref{MHT_BGvsSignal_NoPU} shows the distributions of $\mht$ normalized
to unity for signal (green dotted histogram) and the dominant $V (W,\,Z) +
$jets background (black solid histogram) after all selections except $\mht$
requirement, for the benchmark point with $m_{\tilde{b}} = 500$ GeV,
$m_{\neu{1}} = 495$ GeV, in the case of PU $=0$ for the VBF-like tagged jets
plus $\mht$ study. Based on this distribution, and similar $\mht$
distributions for the VBF-like tagged jets, $\mht$, plus one $b$-tagged jet
study, a shape analysis was performed  with different pileup scenarios. A
local p-value is calculated as the probability under a background only
hypothesis to obtain a value of the test statistic as large as that obtained
with a signal plus background hypothesis. The significance $z$ is then
determined as the value at which the integral of a Gaussian between $z$ and
$\infty$ results in a value equal to the local p-value.

\begin{figure*}[!t]
\centering
\mbox{\includegraphics[width=\columnwidth]{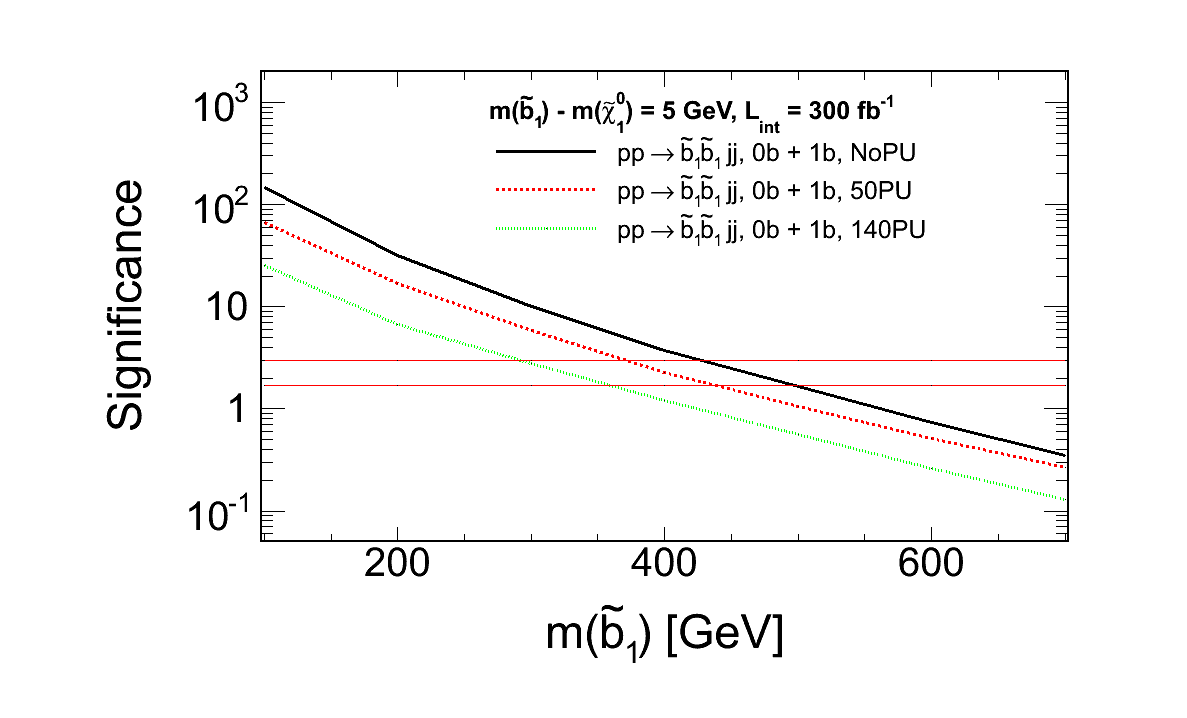}\quad\includegraphics[width=\columnwidth]{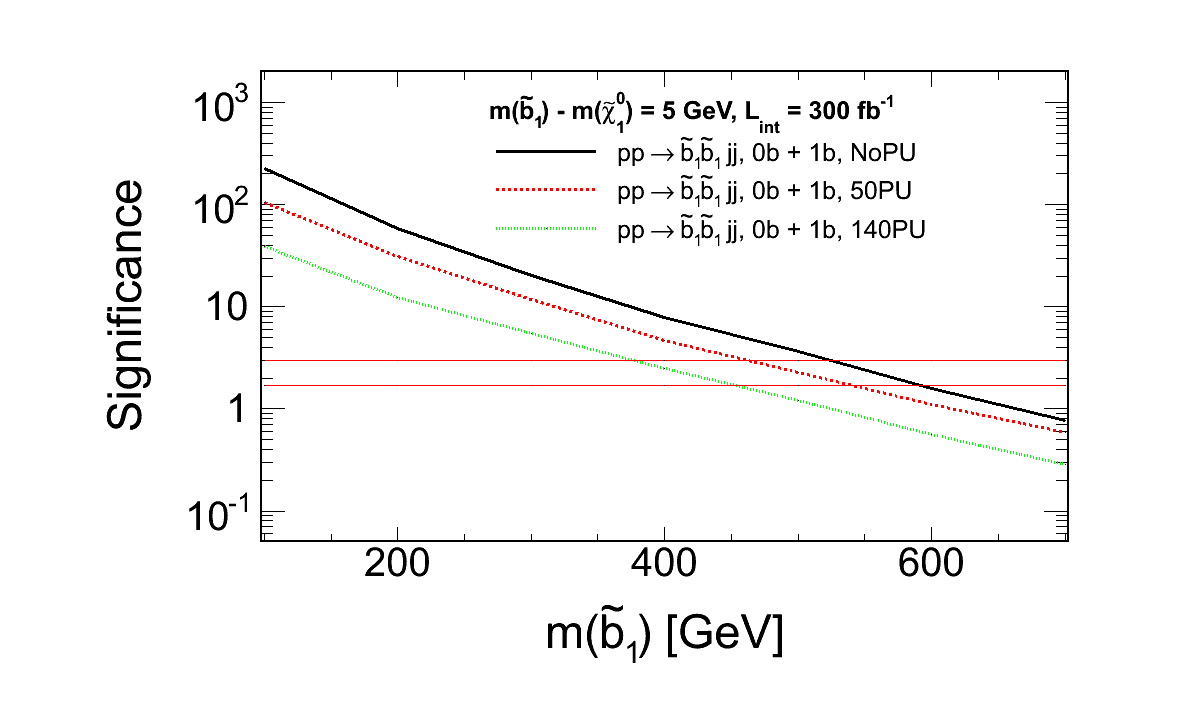}}
\caption{The exclusion reach for the compressed sbottom scenario with
$m_{\tilde{b}} - m_{\neu{1}} \, = \, 5$ GeV. A systematic uncertainty of $5\%$
is assumed throughout. {\bf Left panel:} The significance at $300$ fb$^{-1}$
as a function of $m_{\tilde{b}}$ in the cut and count method, for the joint
studies with and without $b$-tagged jets. From top to bottom, the black solid,
red dashed, and green dotted lines show the cases of PU $ = 0, \, 50, $ and
$140$, respectively. The red solid horizontal lines denote the $3\sigma$ and
$1.69 \sigma$ levels. {\bf Right panel:} The significance at $300$ fb$^{-1}$
as a function of $m_{\tilde{b}}$ with the shape analysis method, for the joint
studies with and without $b$-tagged jets. The legend is identical to the left
panel.}
\label{sigsAllSAndBgfigs}
\end{figure*}

In Figure~\ref{sigsAllSAndBgfigs}, we show the significances of the compressed
scenario with $m_{\tilde{b}} - m_{\neu{1}} \, = \, 5$ GeV, at 300 fb$^{-1}$
with the  cut and count method (left panel) and the shape analysis (right
panel), using joint likelihood to combine the studies with and without
$b$-tagged jets. A systematic uncertainty of $5\%$ is uniformly assumed.  From
top to bottom, the black solid, red dashed, and green dotted curves show the
cases of PU $ = 0, \, 50, $ and $140$, respectively. The red solid horizontal
lines denote the $3\sigma$ and $1.69\sigma$ levels. The shape analysis leads
to a $3\sigma$ exclusion potential of sbottoms with mass up to $530 \, (462)$
GeV with $5\%$ systematics and PU $= 0 \, (50)$.  
(The $95\%$ CL exclusion reach for the PU = 50 case at 300 fb$^{-1}$ is 541 GeV.)
For the most conservative PU
$=140$ case, using the shape (cut and count) analysis, it is possible to probe
compressed sbottoms at the $3\sigma$ level up to $m_{\tilde{b}} = 380 \,(300)$
GeV.

{\it {\bf ISR Monojet study -}}  We now turn to our second analysis, in which
the final state consists of an initial state radiation (ISR) jet, $\mht$, and
at least one $b$-tagged jet. Our analysis mostly follows \cite{Alvarez:2012wf},
which studied this scenario with selections optimized for the 8 TeV LHC. The
event selection is as follows. The $\mht$ asymmetry cut is applied as above
to protect against occasional loss of high $p_{\rm T}$ jets due to pileup subtraction. 
The leading jet is required to be
non $b$-tagged, and have $p_{\rm T} \, > \, 120$ GeV. At least one $b$-tagged
jet with $p_{\rm T} \, > \, 25$ GeV and $|\eta| \, < \, 2.5$ is required. The
leading $b$-tagged jet is required to satisfy $p_{\rm T}(b_1) \, < \, 100$
GeV, with $\Delta \phi (p_{\rm T}(b_1), \met) \, < \, 1.8$.  Leptons are
vetoed. We also require $\mht \, \geq \, 430$ GeV, and the top quark
transverse mass to satisfy $M_{T}^{t}$ $>$ 200 GeV.

\begin{figure}[!htbp]
\centering
\includegraphics[width=3.5in]{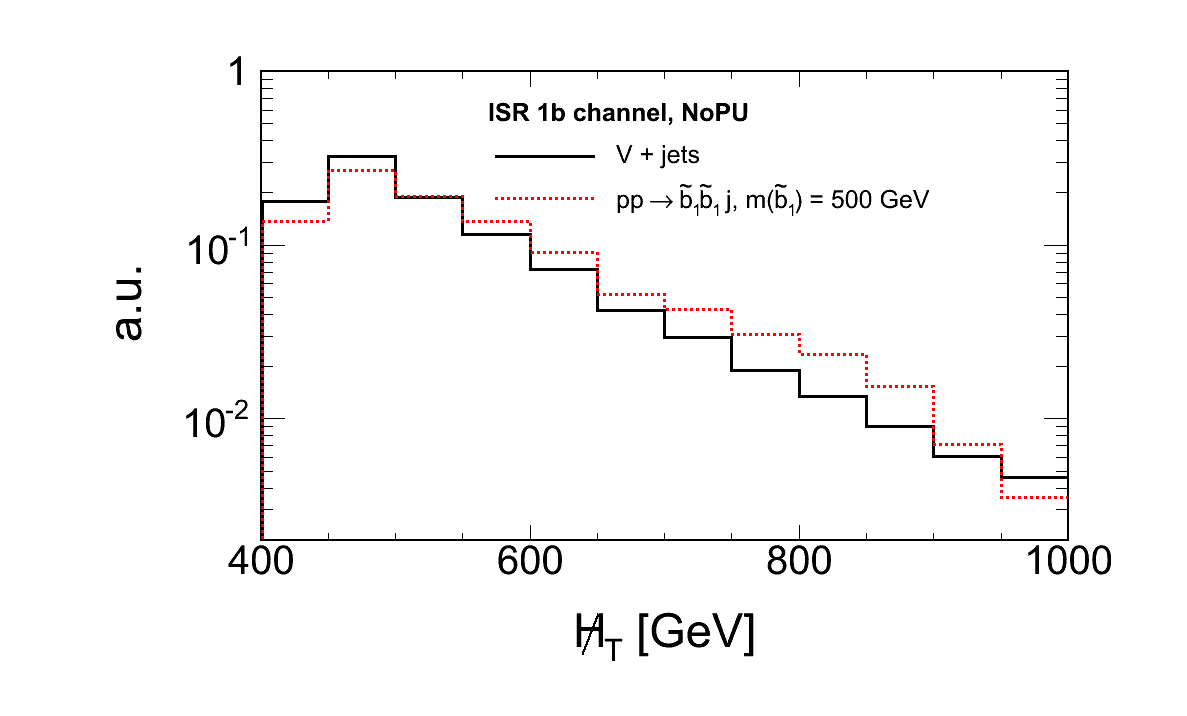}
\caption{{\bf ISR + $b$-tagged jet study:} Distributions of $\mht$ normalized
to unity for signal (red dotted histogram) and dominant $V + $jets background
(black solid histogram) after all selections except $\mht$ requirement, for
the benchmark point with $m_{\tilde{b}} = 500$ GeV, $m_{\neu{1}} = 495$ GeV,
in the case of PU $=0$, for the channel with ISR jet, one $b$-tagged jet, and
$mht$.}
\label{MHT_BGvsSignal_NoPU_ISRb}
\end{figure}

The $\mht$ distribution (after all cuts except $\mht$ cut), normalized to unity, for signal (red dotted histogram) and dominant $V + $jets background (black solid histogram) is shown
in Figure \ref{MHT_BGvsSignal_NoPU_ISRb} for $m_{\tilde{b}} = 500$ GeV and
$m_{\neu{1}} = 495$ GeV, in the case of PU $=0$.

\begin{figure}[!htp]
\centering
\includegraphics[width=3.5in]{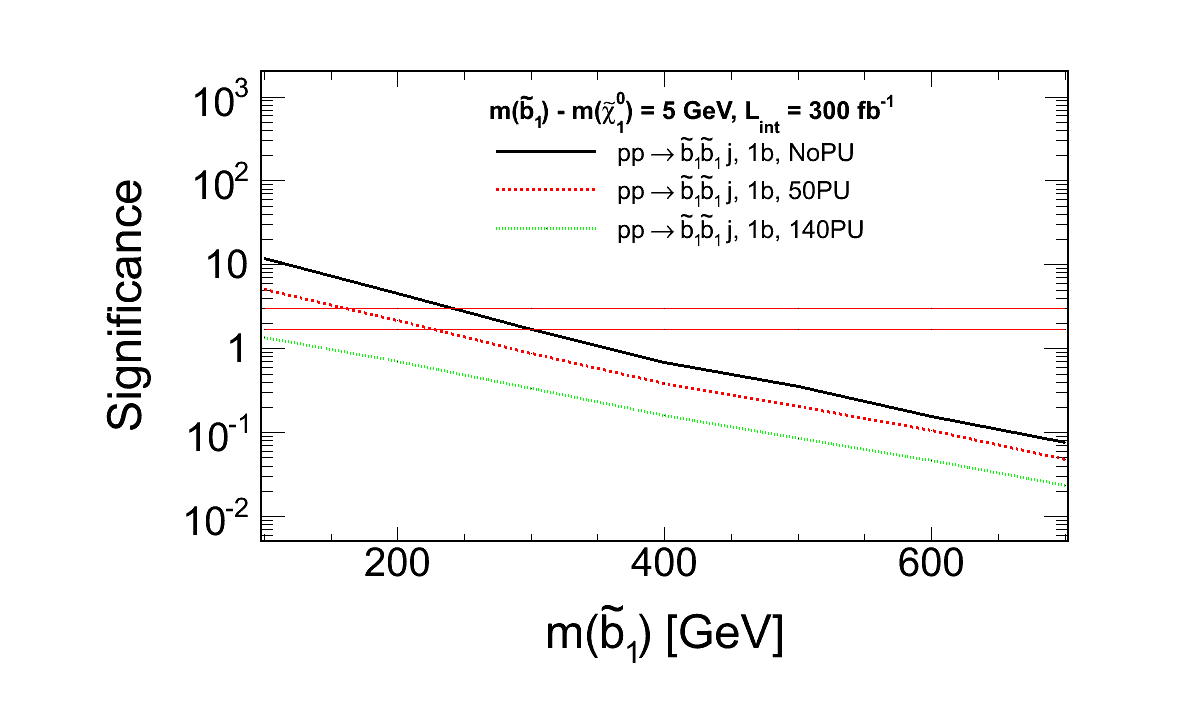}
\caption{{\bf ISR + $b$-tagged jet study:} The exclusion reach is shown for
the compressed sbottom scenario with $m_{\tilde{b}} - m_{\neu{1}} \, = \, 5$
GeV, for the channel with ISR jet, at least one $b$-tagged jet, and $mht$. The
significance at $300$ fb$^{-1}$ as a function of $m_{\tilde{b}}$ after shape
analysis is displayed. A systematic uncertainty of $5\%$ is assumed. From top
to bottom, the black solid, red dashed, and green dotted lines show the cases
of PU $ = 0, \, 50, $ and $140$, respectively. The red solid horizontal lines
denote the $3\sigma$ and $1.69\sigma$ levels.}
\label{ISRbsigplot}
\end{figure}

A shape analysis is performed following the method described above.
This yields the significance plot shown in Figure \ref{ISRbsigplot} for
$m_{\tilde{b}} - m_{\neu{1}} \, = \, 5$ GeV with $300$ fb$^{-1}$ of data. From
top to bottom, the black solid, red dashed, and green dotted lines show the
cases of PU $ = 0, \, 50, $ and $140$, respectively. The red solid horizontal
lines denote the $3\sigma$ and $1.69\sigma$ levels. In the ISR + $b$-tagged
jets study, we found the $3\sigma$ level reach to be $250$ GeV for the PU
$=0$ case.

\begin{figure}[!htp]
\centering
\includegraphics[width=3.5in]{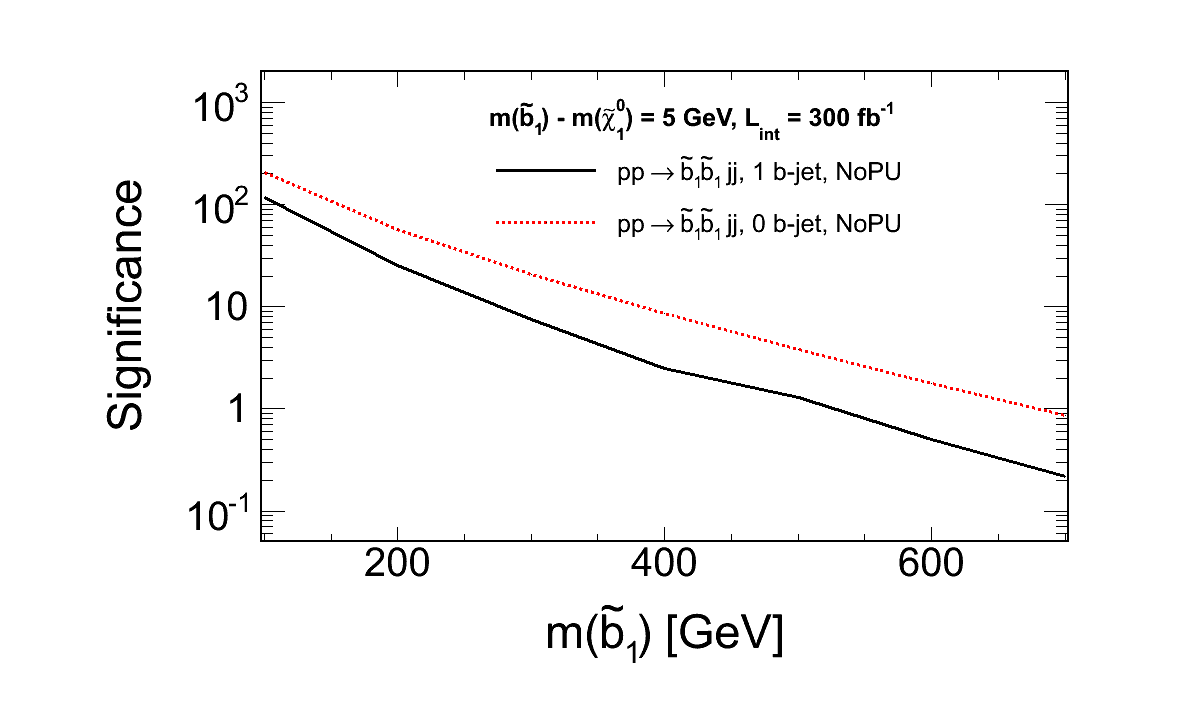}
\caption{{\bf VBF-like tagging jets study:} A comparison of the exclusion
reach with $m_{\tilde{b}} - m_{\neu{1}} \, = \, 5$ GeV, for final states
containing zero $b$-tagged jet (red dotted curve) and one $b$-tagged jet
(black solid line), along with the VBF-like tagged jets and $\mht$. The
significance at $300$ fb$^{-1}$ is given as a function of $m_{\tilde{b}}$
after shape analysis, assuming PU $=0$ and a systematic uncertainty of $5\%$.}
\label{Significance_0bVs1b_Shape_300ifb}
\end{figure}

{\it {\bf Discussion of Snowmass Simulation-}} 
For the study with VBF-like tagged jets, we have performed a joint analysis of
the zero and one $b$-tagged jet final states and displayed the exclusion reach
in the right panel of Figure \ref{sigsAllSAndBgfigs}. Figure
\ref{Significance_0bVs1b_Shape_300ifb} shows the separate significances
in the two final states, after shape analysis, assuming PU $= 0$ and a
systematic uncertainty of $5\%$. It is clear that the performance is dominated
by the zero $b$-tagged jet channel, due to the small mass difference of 5 GeV
between the sbottom and $\neu{1}$. For larger mass difference, it is expected
that the final state containing one $b$-tagged jet would perform better.


In the one $b$-tagged jet analysis, we found that the $b$-tagging efficiency
 at low $p_{\rm T}$ is critical in probing compressed scenarios.
From the Snowmass detector simulation, the efficiency of $b$-tagging reach
60\% at jet $p_{\rm T}$ around 100 GeV \cite{Anderson:2013kxz}, while the CMS
and ATLAS have showed the ability to tag $b$ jet with 60\% efficiency for jets
with $p_{\rm T}$ above 30~GeV~\cite{cmsbjet1, ATLAS:2014jfa}. Thus the
VBF-like tagged jet plus one $b$-tagged jet analysis can be significantly
improved with more efficient and robust $b$-tagging.

Figures~\ref{sigsAllSAndBgfigs} and \ref{ISRbsigplot} shows the degradation in
significance with higher number of pileup for both VBF-like and Monojet
analysis. From \cite{Butler:2020886}, the expected jet performance of the CMS
detector in pileup of 50 is comparable with the performance simulated in the
Snowmass sample~\cite{Anderson:2013kxz}, while the Snowmass samples have much
degraded performance in the 140 pileup condition. With the upgraded CMS and
ATLAS detectors optimized for the high-luminosity LHC condition and
development in the pileup mitigation technique, we are expecting better
physics object performance in the 140 pileup scenario and thus improved reach
of sbottoms in the 140 pileup condition at HL-LHC.

{\it {\bf Conclusions -}}  The main result of this paper is that the boosted
jet topology can provide a feasible strategy to search for compressed bottom
squarks with mass difference of sbottom and lightest neutralino being $5$ GeV.
A shape based analysis is used to estimate the significances. There is
$3\sigma$ exclusion potential up to $530 \, (462)$ GeV for an integrated luminosity of
$300$ fb$^{-1}$ at 14 TeV, with $5\%$ systematics and PU $ = 0(50)$.   
We also performed an ISR + $b$-tagged jet study, and found the exclusion reach to be $250$ GeV for PU $ = 0$.


{\it {\bf Acknowledgements -}}  This work is supported in part by DOE Grant
No. DE-FG02-13ER42020 and DE-FG02-12ER41848 and NSF Award PHY-1206044. T.K.
was also supported in part by Qatar National Research Fund under project NPRP
5 - 464 - 1 - 080. K.S. is supported by NASA Astrophysics Theory Grant
NNH12ZDA001N. Z.W. is supported by National Science Foundation under Grant No.
PHY-1306951. Z.W. would like to thank Richard Cavanaugh for useful
discussions.

\end{document}